\documentclass[aip,reprint]{revtex4-1}

\usepackage{graphicx}
\usepackage{dcolumn}
\usepackage{bm}
\usepackage{color}

\draft 

\begin{document}





\title{Optimizations of force-field parameters for protein systems with
the secondary-structure stability and instability}

\author{Yoshitake Sakae}
\affiliation{Department of Theoretical and Computational Molecular Science, 
Institute for Molecular Science, Okazaki, Aichi 444-8585, Japan}
\affiliation{Department of Physics, Graduate School of Science, 
Nagoya University, Nagoya, Aichi 464-8602, Japan}
\author{Yuko Okamoto}
\affiliation{Department of Physics, Graduate School of Science, 
Nagoya University, Nagoya, Aichi 464-8602, Japan}
\affiliation{Structural Biology Research Center, Graduate School of
Science, Nagoya University, Nagoya, Aichi 464-8602, Japan}
\affiliation{Center for Computational Science, Graduate School of
Engineering, Nagoya University, Nagoya, Aichi 464-8603, Japan}
\affiliation{Information Technology Center, Nagoya University, 
Nagoya, Aichi 464-8601, Japan}



















\begin{abstract}
We propose a novel method for refining force-field parameters of protein systems.
In this method, the agreement of the secondary-structure stability and instability
between the protein conformations obtained by experiments and those obtained
by molecular dynamics simulations
is used as a criterion for the optimization of force-field parameters.
As an example of the applications of the present method,
we refined the force-field parameter set of the AMBER ff99SB force field
by searching the torsion-energy parameter spaces of $\psi$ (N-C$^{\alpha}$-C-N) and $\zeta$ (C$^{\beta}$-C$^{\alpha}$-C-N)
of the backbone dihedral angles.
We then performed folding simulations of $\alpha$-helical and $\beta$-hairpin peptides,
using the optimized force field.
The results showed that the new force-field parameters gave structures more consistent with the
experimental implications than the original AMBER ff99SB force field.
\end{abstract}


\maketitle 



\section{Introduction}
In the field of molecular simulations of protein systems, force fields are widely used.
The force field is represented by a sum of potential-energy terms,
which are given by functions and their parameters based on classical mechanics.
There are several well-known force fields, such as AMBER \cite{parm94, parm96_3, parm99, parm99SB, parm03},
CHARMM \cite{charmm,CMAP}, OPLS \cite{opls1,opls2}, GROMOS \cite{gromos_v2}, and ECEPP \cite{ECEPP}.
When many researchers perform molecular simulations of protein systems, these force fields are utilized extensively.
Generally, the force-field parameters are determined based on experimental results for small molecules
and theoretical results using quantum chemistry calculations of small peptides such as alanine dipeptide.

In a force field, the potential energy is usually composed of the bond-stretching term, bond-bending term,
torsion-energy term, and nonbonded energy term.
In these energy terms, it is known that the torsion-energy term is the most problematic.
For instance, the ff94 \cite{parm94} and ff96 \cite{parm96_3} versions of AMBER differ only in
the backbone-torsion-energy parameters.
Nevertheless, these secondary-structure-forming tendencies are quite different \cite{YSO1,YSO2, SO1, SO2, SO3}.
Therefore, many researchers have studied this energy term and its force-field parameters.
Recently, new force-field parameters of the backbone torsion-energy term about $\phi$ and $\psi$ angles
have been developed, which are, e.g., AMBER ff99SB \cite{parm99SB}, AMBER ff03 \cite{parm03},
CHARMM 22/CMAP \cite{CMAP} and OPLS-AA/L \cite{opls2}.
Newly proposed methods of the force-field refinement also mainly concentrate on the torsion-energy terms.
These modifications of the torsion energy are usually based on
quantum chemistry calculations \cite{Carlos,Duan,IWA,MFB,Kamiya}.
We have also proposed a force-field refinement method in the previous works \cite{SO1,SO2,SO3,SO6,SO7,SO8,SO9}.
These methods are based on the Protein Data Bank (PDB)
and minimize some score functions in the force-field parameter space.
One of the methods consists of minimizing the sum of the square of the force acting on each atom
in the proteins with the structures from the PDB \cite{SO1,SO2,SO3,SO8,SO9}.
Other approaches use the root-mean-square deviation between the original PDB structures and
the corresponding minimized structures as the score functions \cite{SO6,SO7}.

In this article, we introduce a novel optimization method for force-field
parameters, which searches the lowest value of
a score function defined by the sum of numbers of amino acids, which have secondary-structure conformations.
As an example, we applied this approach to AMBER ff99SB force field
and determined new force-field parameters for $\psi$ (N-C$^{\alpha}$-C-N) and $\zeta$ (C$^{\beta}$-C$^{\alpha}$-C-N) angles.
Here, we are using
the Greek letter $\zeta$ for the dihedral angle defined by
C$^{\beta}$-C$^{\alpha}$-C-N.

In Methods section the details of the methodology are given.
In Results and Discussion section the results of applications of the optimization method to
the AMBER ff99SB force field are presented.
The last section is devoted to conclusions.

\section{Methods}
\label{method}
\subsection{Force-field parameters}
The existing force fields for protein systems such as AMBER \cite{parm94, parm96_3, parm99, parm99SB, parm03},
CHARMM \cite{charmm,CMAP},
and OPLS \cite{opls1,opls2} use essentially the same functional forms
for the potential energy $E_{\rm conf}$ except for minor differences.
The conformational potential energy $E_{\rm conf}$ can be written as,
\begin{equation}
E_{\rm conf} = E_{\rm BL} + E_{\rm BA} + E_{\rm torsion} + E_{\rm nonbond}~,
\label{ene_conf}
\end{equation}
where, $E_{\rm BL}$, $E_{\rm BA}$, $E_{\rm torsion}$, and $E_{\rm nonbond}$
represent the bond-stretching term, the
bond-bending term, the torsion-energy term, and the nonbonded energy term, respectively.
Each force field has similar but slightly different parameter values.
For example, the torsion energy is usually given by
\begin{equation}
E_{\rm torsion} = \sum_{{\rm dihedral~angle}~\Phi} \sum_n \frac{V_n}{2} [ 1 + \cos (n \Phi - \gamma_n) ]~,
\label{ene_torsion}
\end{equation}
where the first summation is taken over all dihedral angles $\Phi$ (both in
the backbone and in the side chains),
$n$ is the number of waves, $\gamma_n$ is the phase, and $V_n$ is
the Fourier coefficient.
Namely, the energy term $E_{\rm torsion}$ has $n$, $\gamma_n$, and $V_n$ as force-field parameters.

\subsection{Optimization method of force-field parameters}
We now describe our new method for optimizing the force-field parameters.
In this method, we prepare $M$ protein structures, which are some
experimentally determined conformations.
For these proteins, we perform MD simulations, which start from
the experimental conformations,
by using a trial force field.
We try to perform MD simulations with varied values of force-field parameters.
After that, we estimate the ``$S$'' value defined by the following function
from the trajectories of the $M$ proteins
obtained from  the trial MD simulations:
\begin{equation}
\label{su_num}
S = \sum_{i=1}^{M} \left( \frac{n_i^{\rm S \to U}}{N_i^{\rm S}} + \frac{n_i^{\rm U \to S}}{N_i^{\rm U}}  \right).
\end{equation}
Here,
$n_i^{\rm S \to U}$ is the number of the amino acids in protein $i$ where their
structures in PDB (initial conformation) had some secondary structures
(such as $\alpha$-helix, $3_{10}$-helix, $\pi$-helix, and $\beta$ structures)
but transformed into unstructured, coil structures without any secondary
structures after a short MD simulation.
Likewise,
$n_i^{\rm U \to S}$ is is the number of amino acids in protein $i$
where their structures in PDB had coil structures but transformed
to have some secondary structures after a MD simulation.
$N_i^{\rm S}$ is the total number of amino acids in protein $i$
which have some secondary structures in PDB, and
$N_i^{\rm U}$ is the total number of amino acids in protein $i$
which have coil structures in PDB.

When we calculate the $S$ values for the conformations obtained from MD simulations by using trial force-field parameters,
the parameter set, which yields the minimum $S$ value, is considered to
give the optimized force field.
We remark that we want the number of proteins $M$ and the total number of
time steps of each MD simulation
to be as large as possible in order to reduce the effects of experimental and systematic errors.
In practice, the available computer power decides them.

\section{Results and Discussion}
\label{results}
\subsection{Applications of the optimization method}
\begin{figure}
\begin{center}
\resizebox*{15cm}{!}{\includegraphics{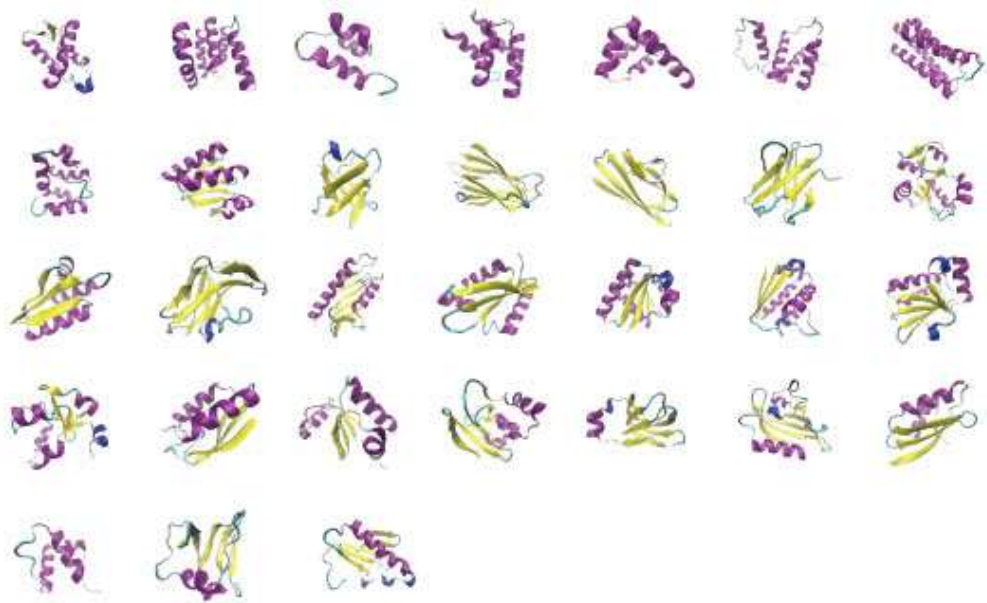}}%
\caption{Structures of 31 proteins from PDB, which were used in 
the optimization of force-field parameters.
The figures were created with VMD \cite{VMD}.}
\label{fig_31strs}
\end{center}
\end{figure}

We present the results of the applications of our optimization method
presented in the previous section to the AMBER ff99SB force field.
We first chose 31 PDB files ($M=31$) with resolution 2.0 \AA~or better,
with sequence similarity of amino acid 30.0 \% or lower,
and with from 40 to 111 residues (the average number of residues is 86.7)
from PDB-REPRDB \cite{REPRDB}.
Namely, the PDB IDs of these 31 proteins are 1LDD, 1HBK, 1Y02, 1I2T, 1U84, 2ERL, 1TQG, 1O82, 1V54, 1XAK,
1GMU, 1O5U, 1NLQ, 1WHO, 1CQY, 1H75, 1GMX, 1IIB, 1VC1, 1AY7, 1KAF, 1KPF, 1BM8, 1MK0, 1EW4, 1OSD, 1VCC,
1OPD, 1CYO, 1CTF, and 1N9L (these structures are shown in Fig.~\ref{fig_31strs}).
We added hydrogen atoms to the PDB coordinates
by using the AMBER11 program package.
We then performed short potential energy minimizations while restraining the coordinates of the heavy atoms.
We used the obtained conformations as the initial structures.
These initial conformations were used as reference experimental structures in the definition of $S$ in Eq.~(\ref{su_num}).
For short MD simulations, the unit time step was set to 2.0 fs and the bonds involving
hydrogen were constrained
by the SHAKE algorithm \cite{SHAKE}.
Each simulation was carried out for 40.0 ps (hence, it consisted of
20,000 MD steps)
by using Langevin dynamics at 300 K.
The nonbonded cutoff of 20 \AA~ were used. As for solvent effects, we used the GB/SA model \cite{gbsa_igb5} included
in the AMBER program package ($igb = 5$).
For all the calculations, we used the AMBER11 program package \cite{amber_prog11}.

\begin{figure}
\begin{center}
\resizebox*{14cm}{!}{\includegraphics{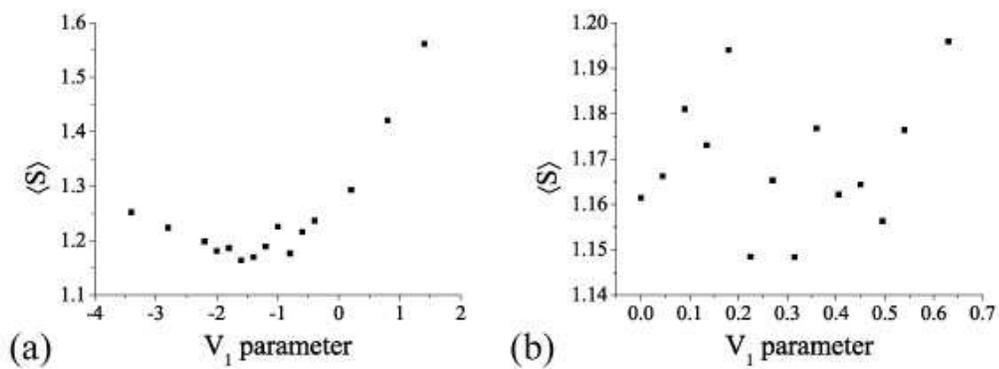}}%
\caption{$S$ values (defined in Eq.~(\ref{su_num})) obtained from
MD simulations of 31 proteins with the force fields which  have
different $V_1$ parameter values for $\zeta$ (C$^{\beta}$-C$^{\alpha}$-C-N)
(a) and $\psi$ (N-C$^{\alpha}$-C-N) (b) angles.}
\label{fig_S_values}
\end{center}
\end{figure}

As the target force-field parameters, we used the parameters $V_1$ of
$\psi$ (N-C$^{\alpha}$-C-N) and $\zeta$ (C$^{\beta}$-C$^{\alpha}$-C-N)
angles for the torsion-energy term in Eq.~(\ref{ene_torsion}).
We performed the simulations by using 14 and 15 different values of the
$V_1$ parameters
of $\psi$ and $\zeta$, respectively, and each simulation with a
fixed set of parameter values was repeated
five times by changing the initial velocities of atoms in
the 31 proteins.
Namely, we calculated $n_i^{\rm S \to U}$ and $n_i^{\rm U \to S}$
in Eq.~(\ref{su_num}) as
the average numbers of $n_i^{\rm S \to U}$ and $n_i^{\rm U \to S}$
of 10 conformations from the last 20.0 ps (within 40.0 ps) for
each simulation (the total number of conformations was thus 50 for
each protein).
The results are shown in Fig.~\ref{fig_S_values}.
We determined the optimized force-field parameters in order of $\zeta$ and $\psi$, by searching the minimum value of $S$
in Fig.~\ref{fig_S_values}.
The optimized $V_1$ value
for $\zeta$ is $-1.60$ (while the original value is 0.20), and
the optimized $V_1$ value
for $\psi$ is 0.31 (while the original value is 0.45).

\subsection{Test simulations of two peptides}
In order to test the validity of the force-field parameters obtained
by our optimization method,
we performed the folding simulations using two peptides, namely,
C-peptide of ribonuclease A and
the C-terminal fragment of the B1 domain of streptococcal protein G, which is sometimes referred to as
G-peptide \cite{gpep3}.
The C-peptide has 13 residues and its amino-acid sequence is
Lys-Glu$^-$-Thr-Ala-Ala-Ala-Lys$^+$-Phe-Glu-Arg$^+$-Gln-His$^+$-Met.
This peptide has been extensively studied by experiments and is known
to form an $\alpha$-helix structure \cite{buzz1,buzz2}.
Because the charges at peptide termini are known to affect
helix stability \cite{buzz1,buzz2},
the N and C termini of the peptide was blocked with acetyl and N-methyl groups, respectively.
The G-peptide has 16 residues and its amino-acid sequence is
Gly-Glu$^-$-Trp-Thr-Tyr-Asp$^-$-Asp$^-$-Ala-Thr-Lys$^+$-Thr-Phe-Thr-Val-Thr-Glu$^-$.
The termini were kept as the usual zwitter ionic states, following the
experimental conditions \cite{gpep3,gpep1,gpep2}.
This peptide is known to form a $\beta$-hairpin structure by experiments
\cite{gpep3,gpep1,gpep2}.

For test simulations, we used replica-exchange molecular dynamics (REMD) \cite{REMD}.
The unit time step was set to 2.0 fs, and the bonds involving hydrogen were constrained by the SHAKE algorithm \cite{SHAKE}.
Each simulation was carried out for 30.0 ns (hence, it consisted of
15,000,000 MD steps) with 32 replicas by using Langevin dynamics.
The replica exchange was tried every 3,000 steps.
The temperature was distributed exponentially:
600, 585, 571, 557, 544, 530, 517, 505, 492, 480, 469, 457, 446, 435, 425, 414, 404, 394, 385, 375, 366,
357, 348, 340, 332, 324, 316, 308, 300, 293, 286, and 279 K.
As for solvent effects, we used the GB/SA model \cite{gbsa_igb5} included in the AMBER program package ($igb = 5$).
These simulations were performed
with different sets of randomly generated initial velocities.

\begin{figure}
\begin{center}
\resizebox*{15cm}{!}{\includegraphics{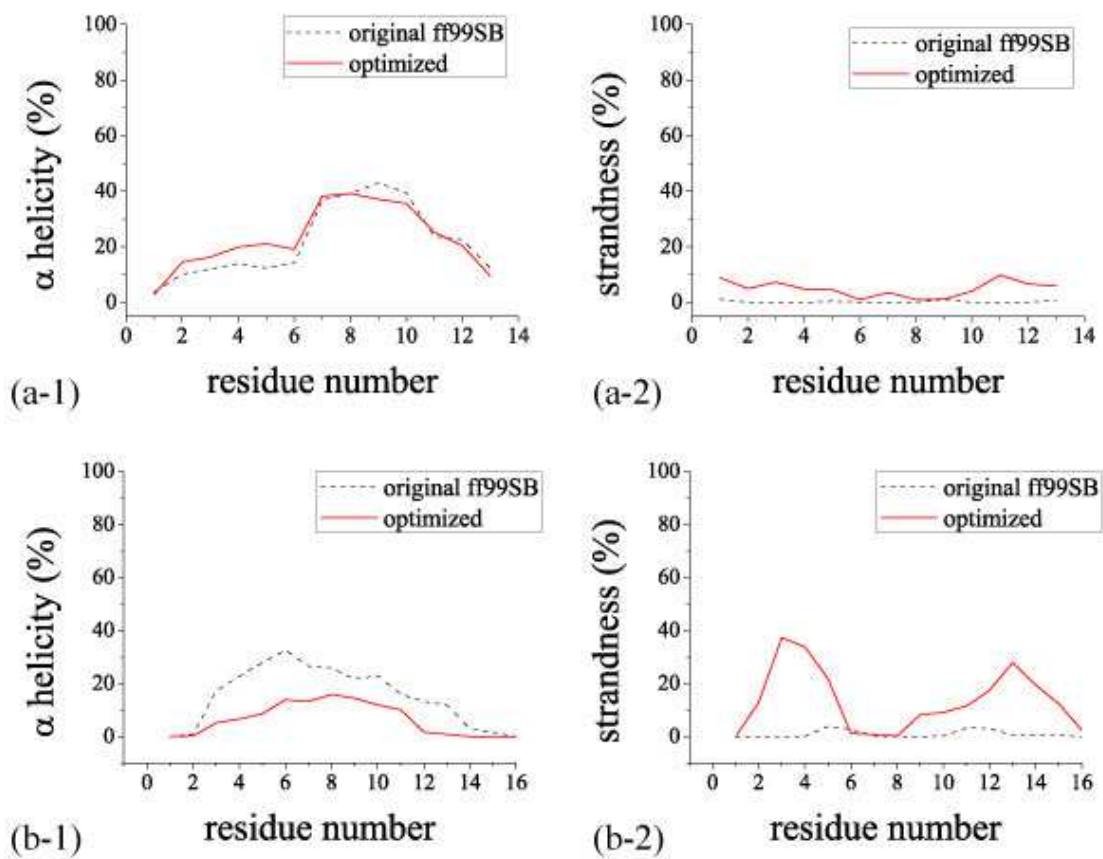}}%
\caption{$\alpha$ helicity (a-1) and strandness (a-2) of C-peptide and $\alpha$ helicity (b-1) and strandness (b-2)
of G-peptide as functions of the residue number. These values were obtained from the REMD simulations at 300 K.
Dotted and normal curves stand for the original AMBER ff99SB and the optimized force field, respectively.}
\label{fig_secondary_alpha_beta}
\end{center}
\end{figure}

In Fig.~\ref{fig_secondary_alpha_beta}, $\alpha$ helicity and strandness of two peptides obtained
from the REMD simulations are shown.
For the original AMBER ff99SB force field, the $\alpha$ helicity is clearly larger than
the strandness in not only C-peptide but also G-peptide.
Namely, the original AMBER ff99SB force field clearly favors $\alpha$-helix structure
and does not favor
$\beta$ structure.
On the other hand, for the optimized force field, in the case of C-peptide, the $\alpha$ helicity is
larger than the strandness, and in the case of G-peptide, the strandness is larger than the $\alpha$ helicity.
We can see that these results obtained from the optimized force field are
in better agreement with
the experimental results than the original force field.

\begin{figure}
\begin{center}
\resizebox*{15cm}{!}{\includegraphics{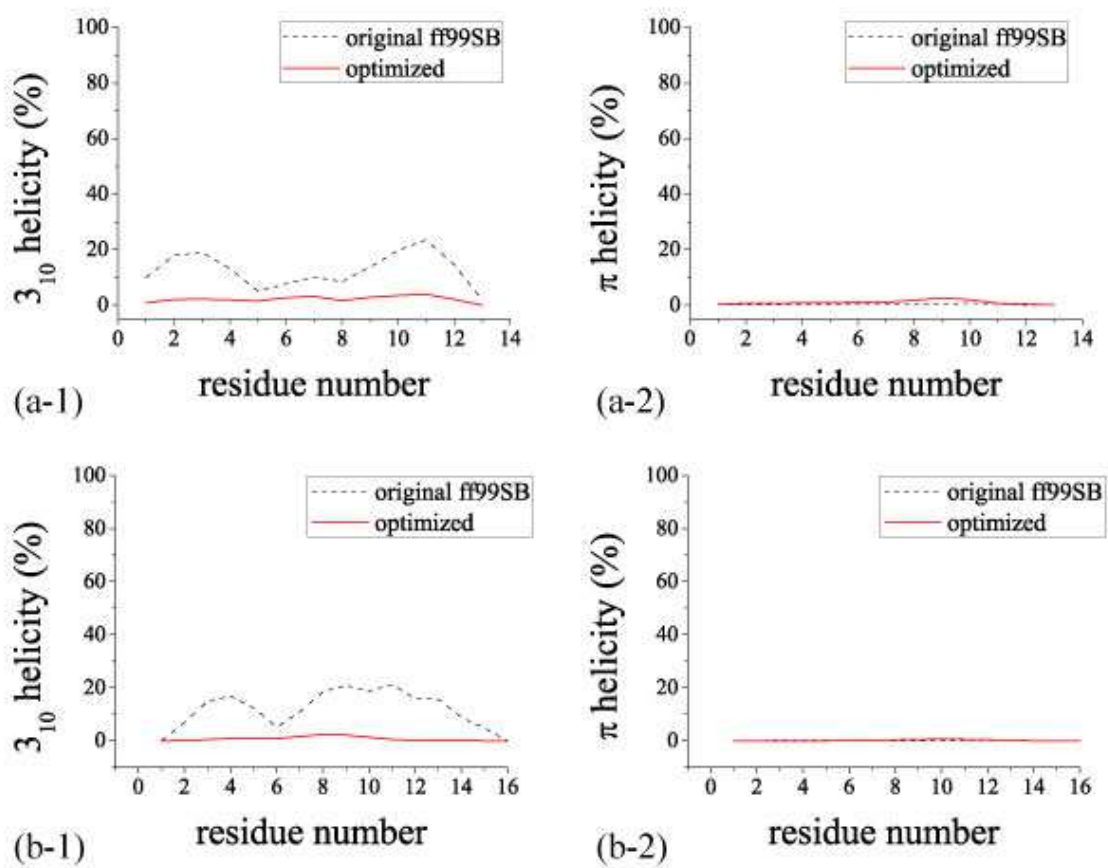}}%
\caption{3$_{10}$ helicity (a-1) and $\pi$ helicity (a-2) of C-peptide and 3$_{10}$ helicity (b-1) and $\pi$ helicity (b-2) 
of G-peptide as functions of the residue number. These values were obtained from the REMD simulations at 300 K.
Dotted and normal curves stand for the original AMBER ff99SB and the optimized force field, respectively.}
\label{fig_secondary_310_pi}
\end{center}
\end{figure}

In Fig.~\ref{fig_secondary_310_pi}, 3$_{10}$ helicity and $\pi$ helicity of two peptides obtained from
the test simulations are shown.
For 3$_{10}$ helicity, the optimized force field does not favor it, while the original force field
favors it to some extent.
$\pi$ helicity has almost no value in the both cases of the original and optimized force fields for the two peptides.
As the both peptides have no 3$_{10}$-helix and $\pi$-helix in the experimental results,
these results of the optimized force field also have better secondary-structure-forming tendencies than those of the original force field.

\begin{figure}
\begin{center}
\resizebox*{15cm}{!}{\includegraphics{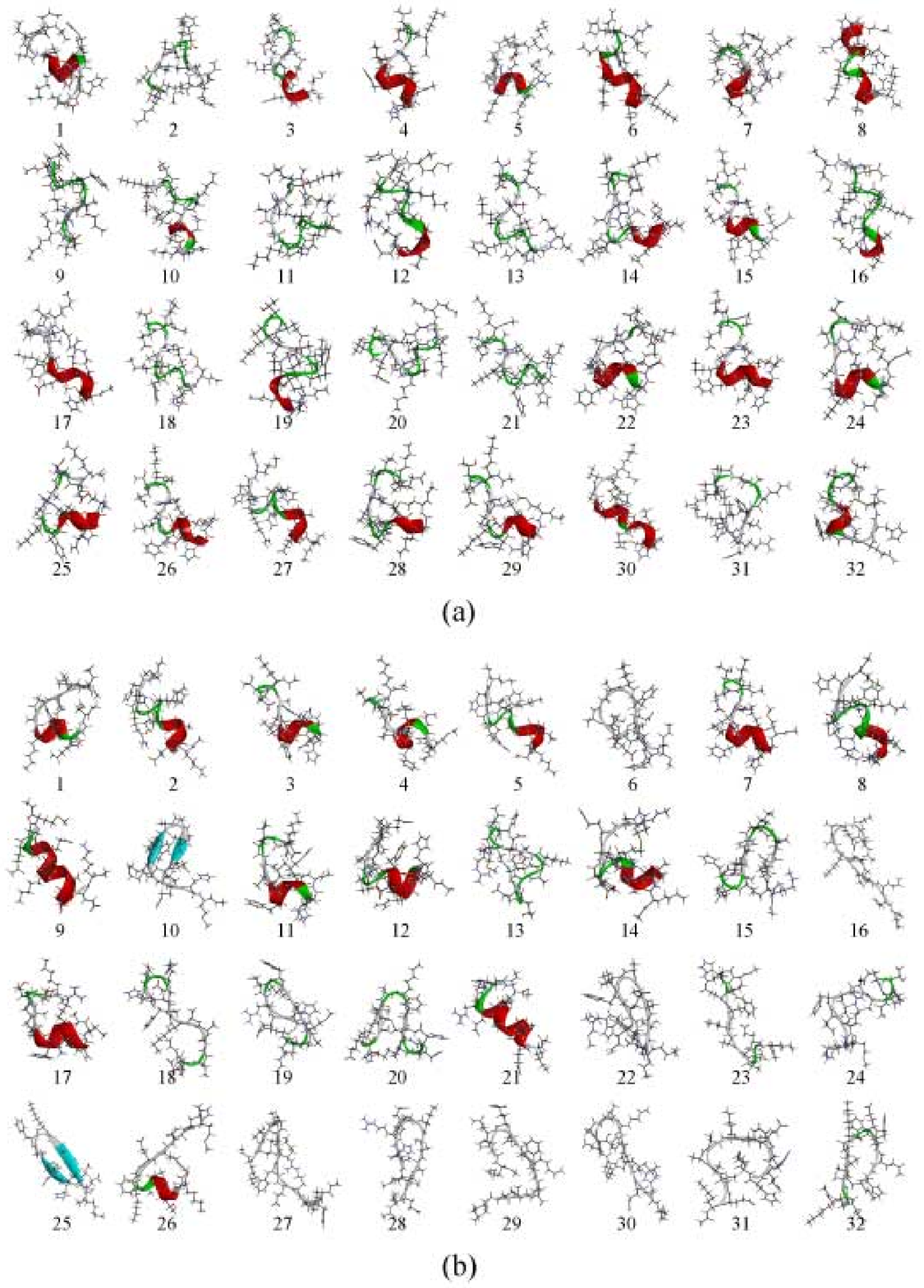}}%
\caption{Lowest-energy conformations of C-peptide obtained for each replica from the REMD simulations. 
(a) and (b) are the results of the original AMBER ff99SB and the optimized force field, respectively.
The conformations are ordered in the increasing order of energy.
The figures were created with DS Visualizer \cite{DSV3}.}
\label{fig_str_cp}
\end{center}
\end{figure}

In Fig.~\ref{fig_str_cp}, the 32 lowest-energy conformations of C-peptide obtained from the REMD simulations for each replica
in the case of the original and optimized force fields are shown.
In the case of the original force field, 27 conformations have helix structures;
15 of them are $\alpha$-helix and 12 are 3$_{10}$-helix.
In the case of the optimized force field, on the other hand, 14 conformations
have helix structures, and all of them are $\alpha$-helix.
Moreover, five conformations are $\beta$ structures.
Both force fields favor helix structures more than $\beta$ structures.
The original force field also has some tendency to favor 3$_{10}$-helix structures,
while the optimized force field does not.
This result is consistent with the results of 3$_{10}$ helicity in
Fig.~\ref{fig_secondary_310_pi}.

\begin{figure}
\begin{center}
\resizebox*{15cm}{!}{\includegraphics{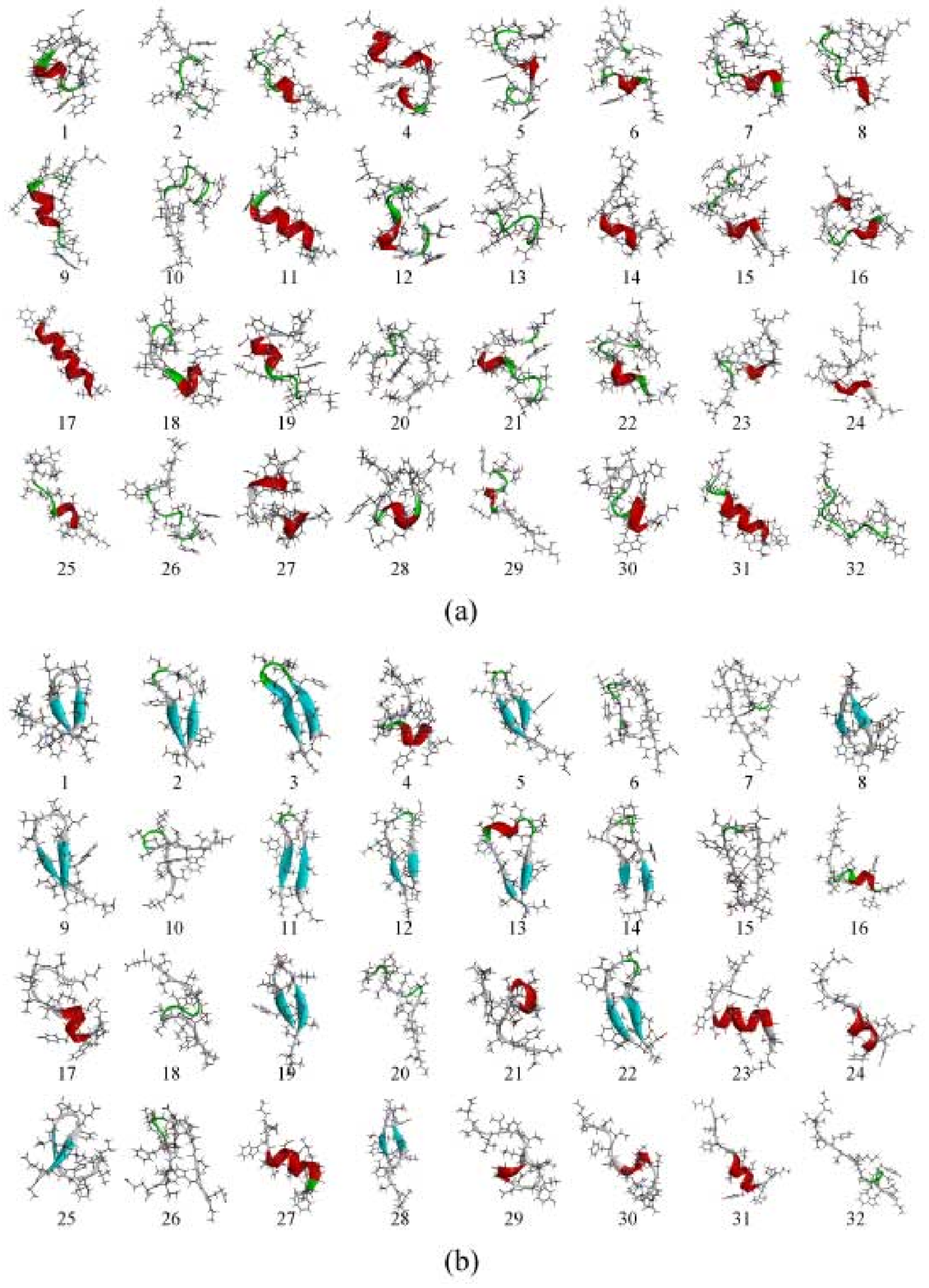}}%
\caption{Lowest-energy conformations of G-peptide obtained for each replica from the REMD simulations. 
(a) and (b) are the results of the original AMBER ff99SB and the optimized force field, respectively.
The conformations are ordered in the increasing order of energy.
The figures were created with DS Visualizer \cite{DSV3}.}
\label{fig_str_gp}
\end{center}
\end{figure}

In Fig.~\ref{fig_str_gp}, the 32 lowest-energy conformations of G-peptide obtained from the REMD simulations for each replica are shown.
As in the case of C-peptide, 27 conformations obtained from the simulations with the original force field have helix structures;
11 of them are $\alpha$-helix, 15 are 3$_{10}$-helix, and one is $\pi$-helix.
In the case of the optimized force field, 11 conformations have helix structure (nine are $\alpha$-helix and two are 3$_{10}$-helix),
and 16 conformations have $\beta$ structures.
We see that the optimized force field slightly favors $\beta$ structure in agreement with
the experimental implications.

\section{Conclusions}
\label{conclusions}
In this work, we proposed a new optimization method of force-field parameters and, as an example, 
optimized some torsion-energy parameters of the ff99SB force field.
This method can optimize force-field parameters using PDB structures.
We applied these optimization method using 31 protein molecules from the Protein Data Bank.
We then performed folding simulations of $\alpha$-helical and $\beta$-hairpin peptides.
We found that the 3$_{10}$ helicity of the optimized force field decreases for both peptides and
that the strandness of the optimized force field increases for $\beta$-hairpin peptide
in comparison with those of the original ff99SB.
The results therefore showed that the optimized force-field parameters gave structures more consistent with the experimental
implications than the original ff99SB force field.
Moreover, the results also showed that the secondary-structure-forming tendencies depend strongly on
only two parameters of backbone-torsion-energy term.
We are now testing the validity of our method by the folding simulations of larger proteins.


\begin{acknowledgments}
The computations were performed on the computers at the Research Center for Computational Science,
Institute for Molecular Science, Information Technology Center, Nagoya University, and Center for 
Computational Sciences, University of Tsukuba.
This work was supported, in part, by
the Grants-in-Aid,
for Scientific Research on Innovative Areas (``Fluctuations and Biological Functions'' ),
and for the Computional Materials Science Initiative
from the Ministry of Education, Culture, Sports, Science and Technology (MEXT), Japan.
\end{acknowledgments}


\begin{mcitethebibliography}{38}
\providecommand*{\natexlab}[1]{#1}
\providecommand*{\mciteSetBstSublistMode}[1]{}
\providecommand*{\mciteSetBstMaxWidthForm}[2]{}
\providecommand*{\mciteBstWouldAddEndPuncttrue}
  {\def\EndOfBibitem{\unskip.}}
\providecommand*{\mciteBstWouldAddEndPunctfalse}
  {\let\EndOfBibitem\relax}
\providecommand*{\mciteSetBstMidEndSepPunct}[3]{}
\providecommand*{\mciteSetBstSublistLabelBeginEnd}[3]{}
\providecommand*{\EndOfBibitem}{}
\mciteSetBstSublistMode{f}
\mciteSetBstMaxWidthForm{subitem}{(\alph{mcitesubitemcount})}
\mciteSetBstSublistLabelBeginEnd{\mcitemaxwidthsubitemform\space}
{\relax}{\relax}

\bibitem[Cornell et~al.(1995)Cornell, Cieplak, Bayly, Gould, Kenneth M.~Merz,
  Ferguson, Spellmeyer, Fox, Caldwell, and Kollman]{parm94}
Cornell,~W.~D.; Cieplak,~P.; Bayly,~C.~I.; Gould,~I.~R.; Kenneth M.~Merz,~J.;
  Ferguson,~D.~M.; Spellmeyer,~D.~C.; Fox,~T.; Caldwell,~J.~W.; Kollman,~P.~A.
  \emph{J. Am. Chem. Soc.} \textbf{1995}, \emph{117}, 5179--5197\relax
\mciteBstWouldAddEndPuncttrue
\mciteSetBstMidEndSepPunct{\mcitedefaultmidpunct}
{\mcitedefaultendpunct}{\mcitedefaultseppunct}\relax
\EndOfBibitem
\bibitem[Kollman et~al.(1997)Kollman, Dixon, Cornell, Fox, Chipot, and
  Pohorille]{parm96_3}
Kollman,~P.~A.; Dixon,~R.; Cornell,~W.; Fox,~T.; Chipot,~C.; Pohorille,~A. The
  development/application of a `minimalist' organic/biochemical molecular
  mechanic force field using a combination of ab initio calculations and
  experimental data. \emph{Computer Simulations of Biological Systems},
  Dordrecht, 1997;
\newblock pp 83--96\relax
\mciteBstWouldAddEndPuncttrue
\mciteSetBstMidEndSepPunct{\mcitedefaultmidpunct}
{\mcitedefaultendpunct}{\mcitedefaultseppunct}\relax
\EndOfBibitem
\bibitem[Wang et~al.(2000)Wang, Cieplak, and Kollman]{parm99}
Wang,~J.; Cieplak,~P.; Kollman,~P.~A. \emph{J. Comput. Chem.} \textbf{2000},
  \emph{21}, 1049--1074\relax
\mciteBstWouldAddEndPuncttrue
\mciteSetBstMidEndSepPunct{\mcitedefaultmidpunct}
{\mcitedefaultendpunct}{\mcitedefaultseppunct}\relax
\EndOfBibitem
\bibitem[Hornak et~al.(2006)Hornak, Abel, Strockbine, Roitberg, and
  Simmerling]{parm99SB}
Hornak,~V.; Abel,~A.,~R.~Okur; Strockbine,~B.; Roitberg,~A.; Simmerling,~C.
  \emph{Proteins} \textbf{2006}, \emph{65}, 712--725\relax
\mciteBstWouldAddEndPuncttrue
\mciteSetBstMidEndSepPunct{\mcitedefaultmidpunct}
{\mcitedefaultendpunct}{\mcitedefaultseppunct}\relax
\EndOfBibitem
\bibitem[Duan et~al.(2003)Duan, Wu, Chowdhury, Lee, Xiong, Zhang, Yang,
  Cieplak, Luo, and Lee]{parm03}
Duan,~Y.; Wu,~C.; Chowdhury,~S.; Lee,~M.~C.; Xiong,~G.; Zhang,~W.; Yang,~R.;
  Cieplak,~P.; Luo,~R.; Lee,~T. \emph{J. Comput. Chem.} \textbf{2003},
  \emph{24}, 1999--2012\relax
\mciteBstWouldAddEndPuncttrue
\mciteSetBstMidEndSepPunct{\mcitedefaultmidpunct}
{\mcitedefaultendpunct}{\mcitedefaultseppunct}\relax
\EndOfBibitem
\bibitem[MacKerell~Jr et~al.(1998)MacKerell~Jr, Bashford, Bellott, Dunbrack,
  Evanseck, Field, Fischer, Gao, Guo, Ha, Joseph-McCarthy, Kuchnir, Kuczera,
  Lau, Mattos, Michnick, Ngo, Nguyen, Prodhom, Reiher, Roux, Schlenkrich,
  Smith, Stote, Straub, Watanabe, Wiorkiewicz-Kuczera, Yin, and
  Karplus]{charmm}
MacKerell~Jr,~A.~D. et~al. \emph{J. Phys. Chem. B} \textbf{1998}, \emph{102},
  3586--3616\relax
\mciteBstWouldAddEndPuncttrue
\mciteSetBstMidEndSepPunct{\mcitedefaultmidpunct}
{\mcitedefaultendpunct}{\mcitedefaultseppunct}\relax
\EndOfBibitem
\bibitem[MacKerell~Jr et~al.(2004)MacKerell~Jr, Feig, and Brooks~III]{CMAP}
MacKerell~Jr,~A.; Feig,~M.; Brooks~III,~C. \emph{J. Comput. Chem.}
  \textbf{2004}, \emph{25}, 1400--1415\relax
\mciteBstWouldAddEndPuncttrue
\mciteSetBstMidEndSepPunct{\mcitedefaultmidpunct}
{\mcitedefaultendpunct}{\mcitedefaultseppunct}\relax
\EndOfBibitem
\bibitem[Jorgensen et~al.(1996)Jorgensen, Maxwell, and Tirado-Rives]{opls1}
Jorgensen,~W.~L.; Maxwell,~D.~S.; Tirado-Rives,~J. \emph{J. Am. Chem. Soc.}
  \textbf{1996}, \emph{118}, 11225--11236\relax
\mciteBstWouldAddEndPuncttrue
\mciteSetBstMidEndSepPunct{\mcitedefaultmidpunct}
{\mcitedefaultendpunct}{\mcitedefaultseppunct}\relax
\EndOfBibitem
\bibitem[Kaminski et~al.(2001)Kaminski, Friesner, Tirado-Rives, and
  Jorgensen]{opls2}
Kaminski,~G.~A.; Friesner,~R.~A.; Tirado-Rives,~J.; Jorgensen,~W.~L. \emph{J.
  Phys. Chem. B} \textbf{2001}, \emph{105}, 6474--6487\relax
\mciteBstWouldAddEndPuncttrue
\mciteSetBstMidEndSepPunct{\mcitedefaultmidpunct}
{\mcitedefaultendpunct}{\mcitedefaultseppunct}\relax
\EndOfBibitem
\bibitem[Gunsteren et~al.(1996)Gunsteren, Billeter, Eising, H{\"u}nenberger,
  Kr{\"u}ger, Mark, Scott, and Tironi]{gromos_v2}
Gunsteren,~W.~F.; Billeter,~S.~R.; Eising,~A.~A.; H{\"u}nenberger,~P.~H.;
  Kr{\"u}ger,~P.; Mark,~A.~E.; Scott,~W. R.~P.; Tironi,~I.~G.
  \emph{Biomolecular Simulation: The GROMOS96 Manual and User Guide};
\newblock Vdf Hochschulverlag AG an der ETH Z{\"u}rich: Z{\"u}rich, 1996\relax
\mciteBstWouldAddEndPuncttrue
\mciteSetBstMidEndSepPunct{\mcitedefaultmidpunct}
{\mcitedefaultendpunct}{\mcitedefaultseppunct}\relax
\EndOfBibitem
\bibitem[N\'emethy et~al.(1992)N\'emethy, Gibson, Palmer, Yoon, Paterlini,
  Zagari, Rumsey, and Scheraga]{ECEPP}
N\'emethy,~G.; Gibson,~K.~D.; Palmer,~K.~A.; Yoon,~C.~N.; Paterlini,~G.;
  Zagari,~A.; Rumsey,~S.; Scheraga,~H.~A. \emph{J. Phys. Chem.} \textbf{1992},
  \emph{96}, 6472--6484\relax
\mciteBstWouldAddEndPuncttrue
\mciteSetBstMidEndSepPunct{\mcitedefaultmidpunct}
{\mcitedefaultendpunct}{\mcitedefaultseppunct}\relax
\EndOfBibitem
\bibitem[Yoda et~al.(2004)Yoda, Sugita, and Okamoto]{YSO1}
Yoda,~T.; Sugita,~Y.; Okamoto,~Y. \emph{Chem. Phys. Lett.} \textbf{2004},
  \emph{386}, 460--467\relax
\mciteBstWouldAddEndPuncttrue
\mciteSetBstMidEndSepPunct{\mcitedefaultmidpunct}
{\mcitedefaultendpunct}{\mcitedefaultseppunct}\relax
\EndOfBibitem
\bibitem[Yoda et~al.(2004)Yoda, Sugita, and Okamoto]{YSO2}
Yoda,~T.; Sugita,~Y.; Okamoto,~Y. \emph{Chem. Phys.} \textbf{2004}, \emph{307},
  269--283\relax
\mciteBstWouldAddEndPuncttrue
\mciteSetBstMidEndSepPunct{\mcitedefaultmidpunct}
{\mcitedefaultendpunct}{\mcitedefaultseppunct}\relax
\EndOfBibitem
\bibitem[Sakae and Okamoto(2003)]{SO1}
Sakae,~Y.; Okamoto,~Y. \emph{Chem. Phys. Lett.} \textbf{2003}, \emph{382},
  626--636\relax
\mciteBstWouldAddEndPuncttrue
\mciteSetBstMidEndSepPunct{\mcitedefaultmidpunct}
{\mcitedefaultendpunct}{\mcitedefaultseppunct}\relax
\EndOfBibitem
\bibitem[Sakae and Okamoto(2004)]{SO2}
Sakae,~Y.; Okamoto,~Y. \emph{J. Theo. Comput. Chem.} \textbf{2004}, \emph{3},
  339--358\relax
\mciteBstWouldAddEndPuncttrue
\mciteSetBstMidEndSepPunct{\mcitedefaultmidpunct}
{\mcitedefaultendpunct}{\mcitedefaultseppunct}\relax
\EndOfBibitem
\bibitem[Sakae and Okamoto(2004)]{SO3}
Sakae,~Y.; Okamoto,~Y. \emph{J. Theo. Comput. Chem.} \textbf{2004}, \emph{3},
  359--378\relax
\mciteBstWouldAddEndPuncttrue
\mciteSetBstMidEndSepPunct{\mcitedefaultmidpunct}
{\mcitedefaultendpunct}{\mcitedefaultseppunct}\relax
\EndOfBibitem
\bibitem[Simmerling et~al.(2002)Simmerling, Strockbine, and Roitberg]{Carlos}
Simmerling,~C.; Strockbine,~B.; Roitberg,~A.~E. \emph{J. Am. Chem. Soc.}
  \textbf{2002}, \emph{124}, 11258--11259\relax
\mciteBstWouldAddEndPuncttrue
\mciteSetBstMidEndSepPunct{\mcitedefaultmidpunct}
{\mcitedefaultendpunct}{\mcitedefaultseppunct}\relax
\EndOfBibitem
\bibitem[Duan et~al.(2003)Duan, Wu, Chowdhury, Lee, Xiong, Zhang, Yang,
  Cieplak, Luo, Lee, Caldwell, Wang, and Kollman]{Duan}
Duan,~Y.; Wu,~C.; Chowdhury,~S.; Lee,~M.~C.; Xiong,~G.; Zhang,~W.; Yang,~R.;
  Cieplak,~P.; Luo,~R.; Lee,~T.; Caldwell,~J.; Wang,~J.; Kollman,~P. \emph{J.
  Comput. Chem.} \textbf{2003}, \emph{24}, 1999--2012\relax
\mciteBstWouldAddEndPuncttrue
\mciteSetBstMidEndSepPunct{\mcitedefaultmidpunct}
{\mcitedefaultendpunct}{\mcitedefaultseppunct}\relax
\EndOfBibitem
\bibitem[Iwaoka and Tomoda(2003)]{IWA}
Iwaoka,~M.; Tomoda,~S. \emph{J. Comput. Chem.} \textbf{2003}, \emph{24},
  1192--1200\relax
\mciteBstWouldAddEndPuncttrue
\mciteSetBstMidEndSepPunct{\mcitedefaultmidpunct}
{\mcitedefaultendpunct}{\mcitedefaultseppunct}\relax
\EndOfBibitem
\bibitem[MacKerell~Jr et~al.(2004)MacKerell~Jr, Feig, and Brooks~III]{MFB}
MacKerell~Jr,~A.~D.; Feig,~M.; Brooks~III,~C.~L. \emph{J. Comput. Chem.}
  \textbf{2004}, \emph{25}, 1400--1415\relax
\mciteBstWouldAddEndPuncttrue
\mciteSetBstMidEndSepPunct{\mcitedefaultmidpunct}
{\mcitedefaultendpunct}{\mcitedefaultseppunct}\relax
\EndOfBibitem
\bibitem[Kamiya et~al.(2005)Kamiya, Watanabe, Ono, and Higo]{Kamiya}
Kamiya,~N.; Watanabe,~Y.; Ono,~S.; Higo,~J. \emph{Chem. Phys. Lett.}
  \textbf{2005}, \emph{401}, 312--317\relax
\mciteBstWouldAddEndPuncttrue
\mciteSetBstMidEndSepPunct{\mcitedefaultmidpunct}
{\mcitedefaultendpunct}{\mcitedefaultseppunct}\relax
\EndOfBibitem
\bibitem[Sakae and Okamoto(2010)]{SO6}
Sakae,~Y.; Okamoto,~Y. \emph{Mol. Sim.} \textbf{2010}, \emph{36},
  159--165\relax
\mciteBstWouldAddEndPuncttrue
\mciteSetBstMidEndSepPunct{\mcitedefaultmidpunct}
{\mcitedefaultendpunct}{\mcitedefaultseppunct}\relax
\EndOfBibitem
\bibitem[Sakae and Okamoto(2010)]{SO7}
Sakae,~Y.; Okamoto,~Y. \emph{Mol. Sim.} \textbf{2010}, \emph{36},
  1148--1156\relax
\mciteBstWouldAddEndPuncttrue
\mciteSetBstMidEndSepPunct{\mcitedefaultmidpunct}
{\mcitedefaultendpunct}{\mcitedefaultseppunct}\relax
\EndOfBibitem
\bibitem[Sakae and Okamoto(2013)]{SO8}
Sakae,~Y.; Okamoto,~Y. \emph{Mol. Sim.} \textbf{2013}, \emph{39}, 85--93\relax
\mciteBstWouldAddEndPuncttrue
\mciteSetBstMidEndSepPunct{\mcitedefaultmidpunct}
{\mcitedefaultendpunct}{\mcitedefaultseppunct}\relax
\EndOfBibitem
\bibitem[Sakae and Okamoto(2013)]{SO9}
Sakae,~Y.; Okamoto,~Y. \emph{J. Chem. Phys.} \textbf{2013}, in press.\relax
\mciteBstWouldAddEndPunctfalse
\mciteSetBstMidEndSepPunct{\mcitedefaultmidpunct}
{}{\mcitedefaultseppunct}\relax
\EndOfBibitem
\bibitem[Noguchi et~al.(1997)Noguchi, Onizuka, Akiyama, and Saito]{REPRDB}
Noguchi,~T.; Onizuka,~K.; Akiyama,~Y.; Saito,~M. PDB-REPRDB: A Database of
  Representative Protein Chains in PDB (Protein Data Bank). \emph{Proc. of the
  Fifth International Conference on Intelligent Systems for Molecular Biology},
  Menlo Park, CA, 1997\relax
\mciteBstWouldAddEndPuncttrue
\mciteSetBstMidEndSepPunct{\mcitedefaultmidpunct}
{\mcitedefaultendpunct}{\mcitedefaultseppunct}\relax
\EndOfBibitem
\bibitem[Ryckaert et~al.(1977)Ryckaert, Ciccotti, and Berendsen]{SHAKE}
Ryckaert,~J.-P.; Ciccotti,~G.; Berendsen,~H. J.~C. \emph{J. Comput. Phys.}
  \textbf{1977}, \emph{23}, 327--341\relax
\mciteBstWouldAddEndPuncttrue
\mciteSetBstMidEndSepPunct{\mcitedefaultmidpunct}
{\mcitedefaultendpunct}{\mcitedefaultseppunct}\relax
\EndOfBibitem
\bibitem[Onufriev et~al.(2004)Onufriev, Bashford, and Case]{gbsa_igb5}
Onufriev,~A.; Bashford,~D.; Case,~D.~A. \emph{Proteins} \textbf{2004},
  \emph{55}, 383--394\relax
\mciteBstWouldAddEndPuncttrue
\mciteSetBstMidEndSepPunct{\mcitedefaultmidpunct}
{\mcitedefaultendpunct}{\mcitedefaultseppunct}\relax
\EndOfBibitem
\bibitem[Case et~al.(2005)Case, Cheatham, Darden, Gohlke, Luo, Merz, Onufriev,
  Simmerling, Wang, and Woods]{amber_prog11}
Case,~D.~A.; Cheatham,~T.; Darden,~T.; Gohlke,~H.; Luo,~R.; Merz,~K.~M.,~Jr.;
  Onufriev,~A.; Simmerling,~C.; Wang,~B.; Woods,~R. \emph{J. Computat. Chem.}
  \textbf{2005}, \emph{26}, 1668--1688\relax
\mciteBstWouldAddEndPuncttrue
\mciteSetBstMidEndSepPunct{\mcitedefaultmidpunct}
{\mcitedefaultendpunct}{\mcitedefaultseppunct}\relax
\EndOfBibitem
\bibitem[Honda et~al.(2000)Honda, Kobayashi, and Munekata]{gpep3}
Honda,~S.; Kobayashi,~N.; Munekata,~E. \emph{J. Mol. Biol.} \textbf{2000},
  \emph{295}, 269--278\relax
\mciteBstWouldAddEndPuncttrue
\mciteSetBstMidEndSepPunct{\mcitedefaultmidpunct}
{\mcitedefaultendpunct}{\mcitedefaultseppunct}\relax
\EndOfBibitem
\bibitem[Shoemaker et~al.(1985)Shoemaker, Kim, Brems, Marqusee, York, Chaiken,
  Stewart, and Baldwin]{buzz1}
Shoemaker,~K.~R.; Kim,~P.~S.; Brems,~D.~N.; Marqusee,~S.; York,~E.~J.;
  Chaiken,~I.~M.; Stewart,~J.~M.; Baldwin,~R.~L. \emph{Proc. Natl. Acad. Sci.
  U.S.A.} \textbf{1985}, \emph{82}, 2349--2353\relax
\mciteBstWouldAddEndPuncttrue
\mciteSetBstMidEndSepPunct{\mcitedefaultmidpunct}
{\mcitedefaultendpunct}{\mcitedefaultseppunct}\relax
\EndOfBibitem
\bibitem[Osterhout~Jr. et~al.(1989)Osterhout~Jr., Baldwin, York, Stewart,
  Dyson, and Wright]{buzz2}
Osterhout~Jr.,~J.~J.; Baldwin,~R.~L.; York,~E.~J.; Stewart,~J.~M.;
  Dyson,~H.~J.; Wright,~P.~E. \emph{Biochemistry} \textbf{1989}, \emph{28},
  7059--7064\relax
\mciteBstWouldAddEndPuncttrue
\mciteSetBstMidEndSepPunct{\mcitedefaultmidpunct}
{\mcitedefaultendpunct}{\mcitedefaultseppunct}\relax
\EndOfBibitem
\bibitem[Blanco et~al.(1994)Blanco, Rivas, and Serrano]{gpep1}
Blanco,~F.~J.; Rivas,~G.; Serrano,~L. \emph{Nature Struct. Biol.}
  \textbf{1994}, \emph{1}, 584--590\relax
\mciteBstWouldAddEndPuncttrue
\mciteSetBstMidEndSepPunct{\mcitedefaultmidpunct}
{\mcitedefaultendpunct}{\mcitedefaultseppunct}\relax
\EndOfBibitem
\bibitem[Kobayashi et~al.(1995)Kobayashi, Honda, Yoshii, Uedaira, and
  Munekata]{gpep2}
Kobayashi,~N.; Honda,~S.; Yoshii,~H.; Uedaira,~H.; Munekata,~E. \emph{FEBS
  Lett.} \textbf{1995}, \emph{366}, 99--103\relax
\mciteBstWouldAddEndPuncttrue
\mciteSetBstMidEndSepPunct{\mcitedefaultmidpunct}
{\mcitedefaultendpunct}{\mcitedefaultseppunct}\relax
\EndOfBibitem
\bibitem[Sugita and Okamoto(1999)]{REMD}
Sugita,~Y.; Okamoto,~Y. \emph{Chem. Phys. Lett.} \textbf{1999}, \emph{314},
  141--151\relax
\mciteBstWouldAddEndPuncttrue
\mciteSetBstMidEndSepPunct{\mcitedefaultmidpunct}
{\mcitedefaultendpunct}{\mcitedefaultseppunct}\relax
\EndOfBibitem
\bibitem[Humphrey et~al.(1996)Humphrey, Dalke, and Schulten]{VMD}
Humphrey,~W.; Dalke,~A.; Schulten,~K. \emph{J. Mol. Grah.} \textbf{1996},
  \emph{14}, 33--38\relax
\mciteBstWouldAddEndPuncttrue
\mciteSetBstMidEndSepPunct{\mcitedefaultmidpunct}
{\mcitedefaultendpunct}{\mcitedefaultseppunct}\relax
\EndOfBibitem
\bibitem[DSV(2007)]{DSV3}
\emph{Discovery Studio Modeling Environment, Release 3.1};
\newblock Accelrys Software Inc.: San Diego, 2007\relax
\mciteBstWouldAddEndPuncttrue
\mciteSetBstMidEndSepPunct{\mcitedefaultmidpunct}
{\mcitedefaultendpunct}{\mcitedefaultseppunct}\relax
\EndOfBibitem
\end{mcitethebibliography}

\providecommand*{\mcitethebibliography}{\thebibliography}
\csname @ifundefined\endcsname{endmcitethebibliography}
{\let\endmcitethebibliography\endthebibliography}{}

\newpage

\end{document}